\documentclass{elsart}
\usepackage{graphicx}
\newcommand{\about}{$\sim$}
\newcommand{\average}[1]{\left< #1 \right>}
\begin{document}
\begin{frontmatter}
\title{ 
 A handy method to monitor outputs \\
 from a pulsed light source \\
 and its application to \\
photomultiplier's rate effect studies }
\author{Y. Takeuchi},
\author{Y. Hemmi\thanksref{DIT}},
\author{H. Kurashige\thanksref{KOBE}},
\author{Y. Matono}, 
\author{K. Murakami},
\author{T. Nomura},
\author{H. Sakamoto}, 
\author{N. Sasao\thanksref{correspond}},
\author{M. Suehiro}
\address{Department of Physics, Kyoto University, Kyoto 606-8502, Japan }
\author{Y. Fukushima},
\author{Y. Ikegami},
\author{T. T. Nakamura}, 
\author{T. Taniguchi}
\address{High Energy Accelerator Research Organization (KEK), 
 Ibaraki 305-0801, Japan }
\author{M. Asai}
\address{Hiroshima Institute of Technology, Hiroshima 731-5193, Japan }
\thanks[correspond]
 {Corresponding author. e-mail: sasao@scphys.kyoto-u.ac.jp}
\thanks[DIT]
 {Present address: {\it Daido Institute of Technology, Aichi 457, Japan}}
\thanks[KOBE]
 {Present address: {\it Kobe University, Hyogo 657-8501, Japan}}

\journal{Nucl. Instr. and Meth. A}

\begin{abstract}
In order to study photomultiplier's short-term gain stability 
 at high counting rate, 
 we constructed an LED pulsed light source and
 its output monitor system. 
For the monitor system, we employed
 a photon counting method
 using a photomultiplier as a monitor photon detector.
It is found that the method offers a simple way
 to monitor outputs from a pulsed light source and that, 
 together with an LED light source, it provides a handy way to
 investigate photomultiplier's rate effects.
\vskip 1em
{\it PACS:\ }29.40.Mc; 85.60.Ha
\end{abstract}
\end{frontmatter}

\section{Introduction}
In high energy physics experiments, photomultipliers are popular devices 
    used as a light-to-charge transducer.
Short-term instability (rate effect) of photomultiplier's gain
    has been a well-known phenomenon~\cite{ref-rate-effect}, 
    which poses one of the major problems to realize good detector
    performance. 
For photomultipliers used in our trigger counter~\cite{ref-E162}, 
    stability was one of the major concerns.
Roughly speaking, gain stability 
    within $\pm 10 \%$ was required up to the counting rate of a few MHz.
More detailed accounts will be given in \S 4.2.

In order to investigate photomultiplier's gain change,
   a stable pulsed light source and/or an
   output light monitor system were needed.
Considering pulse rate involved and handiness,
  the only light source available was LED.
However, LED light outputs might vary substantially at
   a repetition rate of a MHz region.
This motivated us to develop a light monitor system, which
   was suitable to a pulsed light source lit 
   at the repetition rate of MHz or higher.
Needed accuracy for our particular application was a few \%.
We employed a photon counting method for this purpose.
We describe the results of studies on the LED light source and
  its monitor system in the following sections.
This report is organized as follows: in \S 2 the principle of the
    photon counting method is described.
The experimental setup and the results of test measurements 
    are shown in \S 3 and \S 4, respectively.
The section 5 summarizes our studies.

\section{Principle of the Method}
\begin{figure*}[tbp]
 \begin{center}
  \includegraphics[height=170bp, clip]{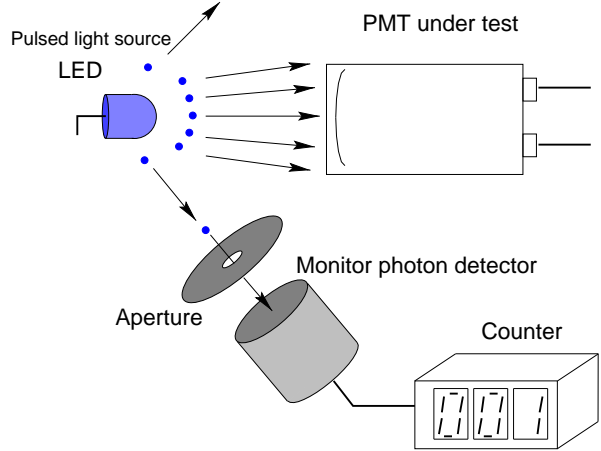}
  \caption{Schematic diagram showing the principle.}
  \label{fig-principle}
 \end{center}
\end{figure*}
Fig.\ref{fig-principle} shows a schematic diagram 
  which illustrates our method.
A photomultiplier under test, placed in front of a pulsed 
 LED, receives most of the light output.
We sample a very small portion of the lights and inject it to
 a monitor photon detector. 
Let $\eta_{samp}$ denotes the sampling fraction of photons; 
 it mainly depends on geometrical factors such
 as the distance between the light source and detector,
 and, if exist, an  
 aperture and attenuation filters between them.
We regard this fraction to be practically constant 
 during the course of a measurement.
The expected number of photons per pulse
 detected by the monitor detector is given by 
 $ \average{n} = \average{N_{LED}} \cdot \eta_{samp} \cdot \eta_{det}$,
 where $ \average{N_{LED}} $ 
 represents the average number of photons per pulse emitted by LED, 
 and $\eta_{det}$ the monitor's detection efficiency.
The probability distribution for $\, n \,$ is given
 by the Poisson distribution. 
In this method we adjust $ \average{n}$,
 the average number of photons per pulse,  
  to be much less than unity.
This can be done at will by changing, for example, 
  aperture size or attenuation filters.
Since the probability to detect one or more photons per pulse
 is given by
     \begin{eqnarray*}
     P(n \ge 1)=1-e^{-\average{n}},
     \end{eqnarray*}
 $ \average{n} $ can be represented by
     \begin{eqnarray*}
     \average{n}=- \log \{ 1 - P(n \ge 1) \}.
     \end{eqnarray*}
We can monitor $\average{n}$ by measuring $P(n \ge 1)$,
 and thus the average LED light output assuming 
 $\eta_{samp} \cdot \eta_{det}$ to be constant.
Here an important feature required for the monitor photon detector 
  is capability of discriminating 
  the single photon signal from background noise.
In our actual setup we used a photomultiplier as a monitor detector.
As described in the following section in detail,
  we could distinguish clearly a single photoelectron 
  peak from a pedestal\cite{ref-photoelectron}.
We measured the counts in which outputs from the monitor
  photomultiplier exceeded some prescribed level set between
  the pedestal and single photoelectron peak.
Then $P(n \ge 1)$ was given by the counts divided by the 
  total number of pulses which triggered the LED light source.

The principal advantages of the method are the followings:
\begin{description}
\item[{\rm (i)}]
As long as the single photon signal can be discriminated from background 
  noise, small gain variation of a monitor photon detector itself
  has almost no effect on $P(n \ge 1)$ and
  thus $\average{n}$.
This is the most important feature in this method.
By contrast, if $\average{n}$ is much bigger than one and
  the monitor photon detector measures the 
  light output every pulse,
  it is not possible to distinguish the change in the LED
  output itself from the gain variation in the
  monitor detector.
\item[{\rm (ii)}]
If the peak corresponding to the single photon 
  can be observed, the gain change may be monitored by
  measuring its peak position.
This feature is helpful to demonstrate reliability
  in the monitor detector.
\item[{\rm (iii)}]
The monitor detector must be able to discriminate the
  single photon from background noise, as mentioned above.
However, it is not necessary to resolve single photon from
  two (or more) photons since the measured quantity is $P(n \ge 1)$.
This lessens requirement for the monitor detector.
\item[{\rm (iv)}]
The actual counting rate for the monitor detector 
 can be set low  by adjusting 
 $\average{n}$ to  be much less than unity.  
We note that monitor detectors are usually more stable
 at lower counting rates for a fixed gain.
\end{description}

Disadvantages of the method, on the other hand,
 are that it monitors not instantaneous but average light outputs,
 and that it takes rather long time to obtain enough statistical accuracy. 
For example, when an LED is lit at 10 kHz and $\average{n}$ is
 $\sim 0.01$, then it takes 100 sec to obtain $10^{4}$ counts,
 the number of counts needed to reach the statistical accuracy of $1 \, \%$.
It is expected that main source of the systematic errors for the
  method stems from various backgrounds to the monitor photon detector.
It turns out that thermal noises and after-pulses
  are the two major backgrounds 
  when a photomultiplier is used as a monitor photon detector.
We thus studied these backgrounds carefully (see below for the detail).

\section{Experimental Setup}
\begin{figure*}[tbp]
 \begin{center}
  \includegraphics[height=200bp, clip]{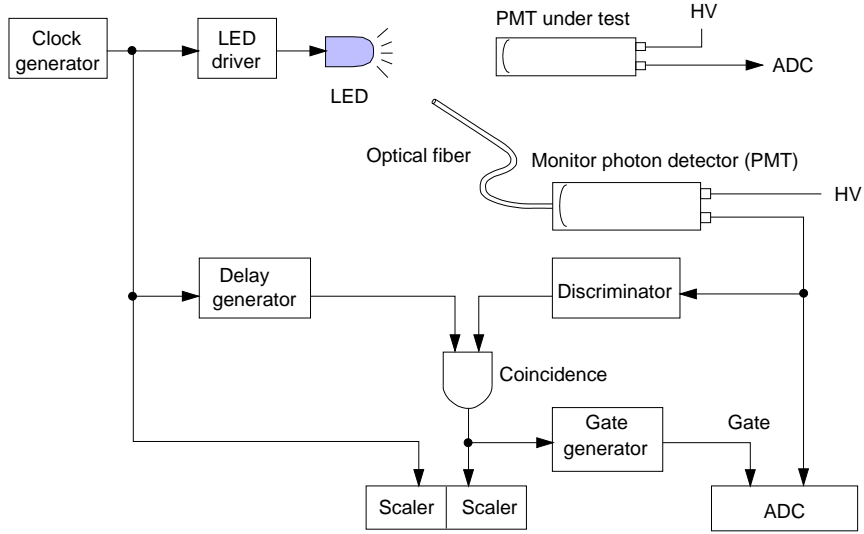}
  \caption{Schematics of the monitor system.}
  \label{fig-setup}
 \end{center}
\end{figure*}

As shown in Fig.\ref{fig-setup}, the system 
 consisted of a light source, 
 a quartz fiber for photon sampling, a monitor photon detector, and
 a trigger and read-out electronic system.
A photomultiplier subject to rate effect studies 
 was placed in front of the light source.
A brief description of each component is given below.
\subsection{Pulsed light source (LED)}
We used a `blue' LED~\cite{ref-blueLED} as a light source. 
For the present application, it was found advantageous to
  use blue in two respects.
First, the tail of its light output was 
  substantially shorter(\about 20 nsec) 
  than that of a `green' 
  LED (\about 50 nsec)~\cite{ref-greenLED}~\cite{note-LED-property}.
Secondly, the emission spectrum of the blue LED 
  resembled more to that of the scintillator we used;
  a desirable property for the photomultiplier gain test.

\begin{figure*}[htbp]
 \begin{center}
  \includegraphics[height=200bp, clip]{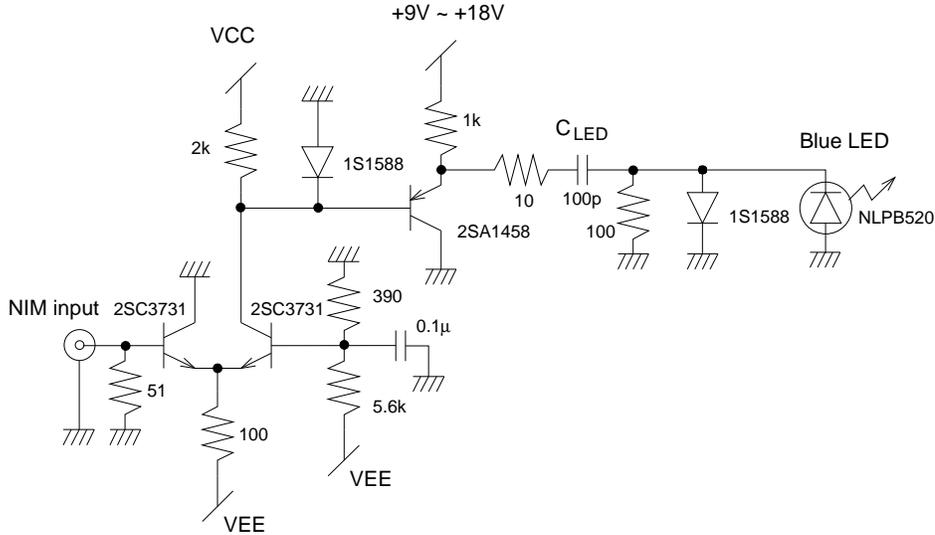}
  \caption{LED driver circuit.}
  \label{fig-circuit}
 \end{center}
\end{figure*}

Fig.~\ref{fig-circuit} shows the LED driver circuit.
The circuit provided a constant charge to the LED
  stored in the capacitor $C_{LED}$.
The discharge was triggered by a differential switch,
  which in turn initiated by an external NIM pulse.
After the discharge, the capacitor $C_{LED}$ was recharged by 
  an external power supply with the charge-up time constant
  $\tau$ of 0.1 $\mu$sec.
\subsection{Trigger and read-out system}
A variable frequency NIM clock generator
  was used to produce a master pulse.
Its output  was fed to the LED driver and to a scaler.
Output signals from the monitor photon detector,  
  a photomultiplier for the present case, 
  were discriminated and fed into
  a coincidence circuit.
We set the discriminator threshold level at about 1/4
 of the single photoelectron peak.
The coincidence signal of the discriminator output and
 the master clock produced a 
 60-nsec-long gate to a charge sensitive ADC, which integrated
 the raw signal from the monitor photon detector.
The gate signal was also counted by another scaler.
Data from the ADC and scalers were read by a computer via a CAMAC
 system.

\subsection{Monitor photon detector}
As stated, we used a photomultiplier as a monitor photon detector. 
Selection of an actual photomultiplier was made by considering 
  (i) single-photoelectron resolution, (ii) thermal noise rate,
  and (iii) after-pulse rate.
We tested the following types of photomultipliers;
  Hamamatsu R329, R1332, R2165 and R3234~\cite{ref-Hamamatsu}.
It was found that backgrounds
  due to the after-pulse 
  depended strongly upon photomultiplier types, 
  and that, for some of them, they were the source of the 
  most serious systematic errors.
We finally chose R3234 from those listed above with an
 emphasis on this point.
\begin{figure*}[tbp]
 \begin{center}
  \includegraphics[height=160bp, clip]{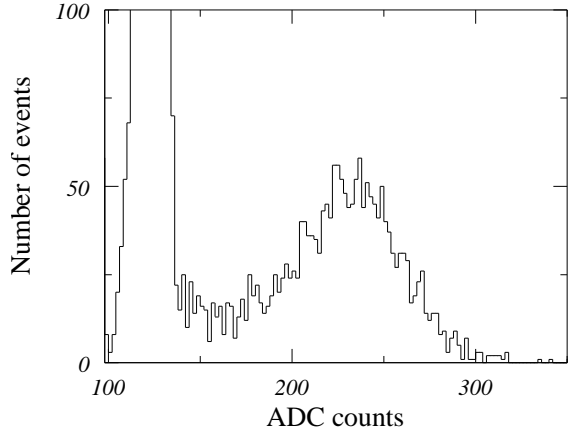}
  \caption{Pulse height spectrum obtained with R3234.}
  \label{fig-single}
 \end{center}
\end{figure*}
\paragraph{Single-photoelectron resolution}
Fig.\ref{fig-single} shows the pulse height spectrum obtained
  with the photomultiplier actually used;
  the left peak  (scaled off) corresponds to a pedestal
  and the right to the single photoelectron.
We defined a signal count as an event
  above a software cut placed at the bottom of the valley 
  on this spectrum.
The actual cut position was 0.4 in units of the single photoelectron,
  i.e. the ADC counts between the pedestal and single photoelectron 
  peak.
It was confirmed that variation of the cut position
  within a reasonable range from the nominal value 
  resulted a negligible change in the final results~\cite{ref-cutposition}.
\paragraph{Thermal noise}
Thermal (random) noises may contribute to a systematic error. 
We measured the noise rate 
 and found it to be $\sim$ 400 Hz at 15 $^\circ \rm C$. 
The background count per pulse is then
 \about 400 Hz$\times $60 nsec (ADC gate width) $= 2.4 \times 10^{-5}$.
This should be compared with an average signal rate of $\average{n}$.  
The background is thus severe at small $\average{n}$;
 however, it is possible to adjust $\eta_{samp}$ so that  $\average{n}$
is much larger than the noise contribution.
In our actual measurements ( see \S 4.2 for an example ),
  the lowest value for $\average{n}$ was chosen to be 0.006.
Thus the error due to this noise is negligible
  ($< 1 \%  $)~\cite{note-Thermal-exception}.
\begin{figure*}[tbp]
 \begin{center}
  \includegraphics[scale=1.0, clip]{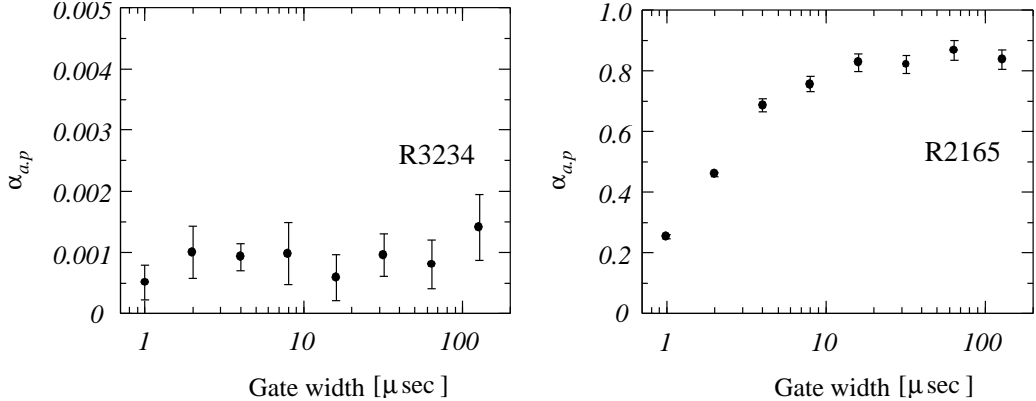}
  \caption{After-pulse probability $\alpha_{a.p}$ for R3234 (left) 
    and R2165 (right).}
  \label{fig-afR3234}
 \end{center}
\end{figure*}
\paragraph{After-pulse}
An after-pulse is a spurious pulse induced in a photomultiplier by  
  previous pulses~\cite{ref-after-pulse}.
It is induced by positive ion hits on a photocathode 
 which is produced by collisions between electrons and residual
 gas molecules in the tube. 
Since original electron currents are initiated by input light,
 the after-pulse has time and rate correlation with the input light.
Let's denote by $\alpha_{a.p}$ the average number of after-pulses 
  per input light which emits single photoelectron. 
Then, in the worst case, namely when the after-pulse happens to
  have a complete time correlation with 
  the following signal pulse, $\average{n}$ would increase
  to $\average{n}(1+\alpha_{a.p})$. 
Thus the only way to reduce this error is to choose a 
  photomultiplier with small $\alpha_{a.p}$.

In order to find an appropriate photomultiplier,
  we measured this quantity  $\alpha_{a.p}$.
The measurement was done with the same setup
  shown in Fig.\ref{fig-setup} with one minor change;
  the delay generator 
  started the gate pulse about 200
  nsec after the LED light pulse. 
The gate width determined the
  time interval to look for the after-pulses.
The results are shown in Fig.\ref{fig-afR3234}
 for the photomultiplier we selected (R3234), together with 
 another type of 2-inch photomultiplier (R2165) for comparison.
In the figures, the abscissa represents the integration period
 (the gate width) while the ordinate represents the after-pulse
 probability $\alpha_{a.p}$.
As can be seen, the integrated counts saturate
 around the gate width of 10 $\mu$sec.
 From the results above and similar measurements
 for the other types of photomultipliers listed above, 
 we concluded that the integration time of 128 $\mu$sec 
 was long enough to
 detect practically all the after-pulses.
The selected photomultiplier R3234 has particularly small 
 value ($\alpha_{a.p} \sim 0.001)$~\cite{note-ap-pmt}.
We thus expect the error due to this background is also negligible.

\section{ Results of Test Measurements}
\subsection{Cross-check measurement}
It is difficult to determine the absolute 
 accuracy of this method  experimentally since there is no `ideal' 
 light source to calibrate with. 
Nevertheless we wanted to obtain a crude `estimate' of its accuracy,
 and thus compared it with one other method.

\begin{figure*}[htbp]
 \begin{center}
  \includegraphics[height=200bp, clip]{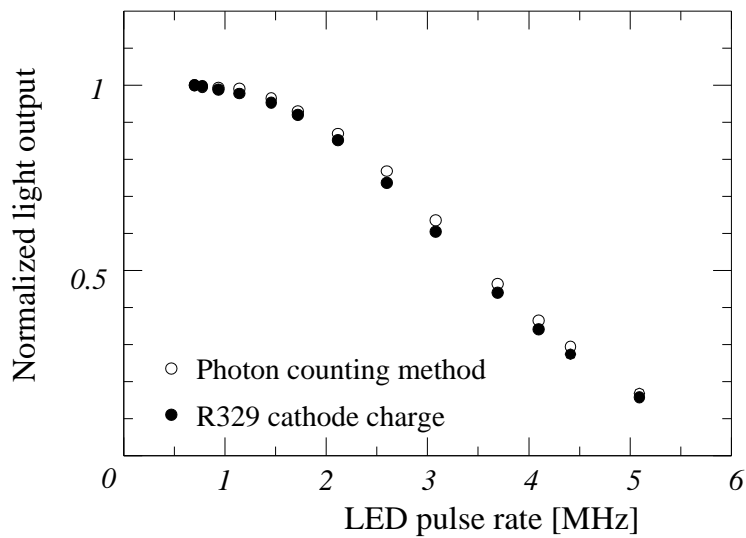}
  \caption{Normalized LED light outputs measured 
  by R329 and by the photon counting method 
  as a function of pulse rate.}
  \label{fig-cathode}
 \end{center}
\end{figure*}

In place of a test photomultiplier, we set an R329
 photomultiplier operated in a diode mode.
This was accomplished by keeping the cathode 
 at -300 V while all the other dynodes grounded.
The average cathode current was measured by 
 an amplifier and a current monitor. 
Since there was no electron multiplication involved,
 the output current was expected to be proportional to the 
 input light even at high rate.
At low rate, however, the output was dominated by electronic noise,
 and the measurement became less accurate.
Actually we measured the output current
 produced by the LED light pulse at the repetition rate between 
 0.7 MHz and 5 MHz~\cite{note-LED-drop}.
The measured values of the current from R329 were converted
 to the charge per pulse and then normalized to that at 0.7 MHz.
The resultant quantities, namely the normalized LED light outputs per pulse
 as a function of pulse rate, are displayed in Fig.\ref{fig-cathode},
 together with the corresponding quantities obtained 
 with the photon counting method.
As seen, they agree fairly well with each other up to 5 MHz.  
The maximum deviation is found to be about $7 \% $.
The origin of the discrepancy is not clear at present~\cite{note-Thermal-systematic}.

\subsection{Example of the stability measurement}
In this subsection, we show an example of the 
 gain stability measurement performed with this system.
The photomultipliers under test were used in the trigger
 counter in our experiment~\cite{ref-E162}.
Their short-term stability was one of the major concerns 
  because of the following reasons.
The trigger counter was composed of a set of plastic scintillator slabs
    and was installed in an intense neutral $K$ beam.
Scintillation lights from the counter were read by photomultipliers 
   attached at the both ends of the scintillators.
We chose Hamamatsu R1398\cite{ref-Hamamatsu}, a photomultiplier with a 
  bialkali photocathode which had a spectral response well matched
  with an emission spectrum of the scintillator, and a linear focused dynode chain
  which provided a fast rise time and a good pulse linearity.
These properties, together with its cathode diameter 
   (1-1/8$^{\prime \prime}$), were well suited to our
   application.
We used an AC-coupled preamplifier and base-line restorer as
   a part of the read-out circuit.
The preamplifier (with $\sim$ 30 db gain)
  helped to reduce photomultiplier's average anode current while 
  the base-line restorer compensated base-line shifts at high counting rate.
If the photomultiplier gains were to be set high to compensate
    possible gain drop at high counting rate, hit rates 
    would increase by background particles 
    such as neutrons and gammas in the beam.
For our experiment ($K_{L}$ rare decay), 
    these background hits should be avoided as much 
    as possible to reduce trigger rates and to ensure high
    reconstruction efficiency in off-line analysis.
In addition a large pulse would tend to cause a longer
    dead time for a preamplifier due to saturation,
    making the counter inefficient.
As a consequence the photomultiplier gain should be kept 
    as low as practical while maintaining $\sim 100 \% $ efficiency 
    for the minimum ionizing particles. This demanded good 
    gain stability (say relative gain change within $\pm 10\%$ ) 
    at the expected  highest counting rate
    ( i.e., $\sim 4$ MHz for each photomultiplier).

\begin{figure*}[htbp]
 \begin{center}
  \includegraphics[height=200bp, clip]{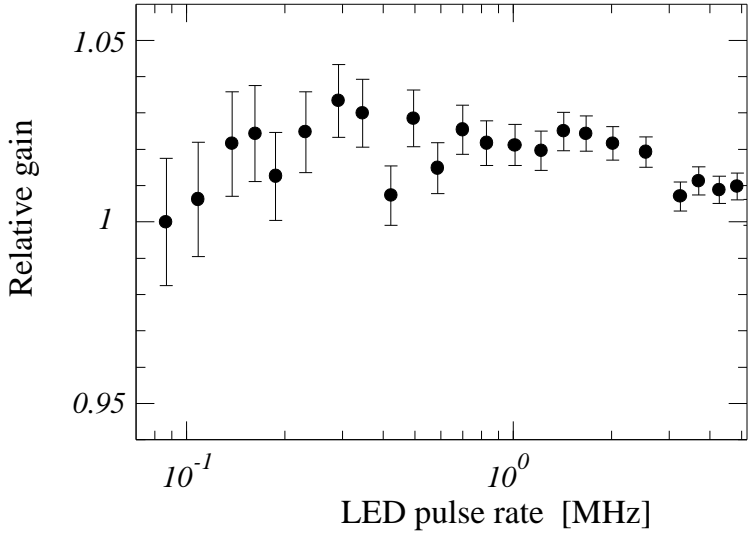}
  \caption{Gain stability of R1398 as a function of pulse rate.}
  \label{fig-R1398}
 \end{center}
\end{figure*}

Fig.\ref{fig-R1398} shows the result for the R1398 type photomultiplier.
It shows the R1398 output divided by $\average{n}$ 
 as a function of the LED pulse rate; 
 the ratio is then rescaled to 1 at 86 kHz. 
In the measurement, the LED light intensity on R1398 was adjusted
  to give approximately 100 photoelectrons per pulse independent of 
  the pulse rate, which
  was approximately equal to the average
  scintillator light output produced by minimum ionizing particles
  passing through our trigger counter.
We accomplished this by pulsing 7 identical LEDs in turn,
  thus keeping the effective pulse rate for any individual LED
  less than 1 MHz~\cite{note-7LEDs}.
The maximum deviation of the normalized R1398 output 
  from the unity is found to be about 3\% in the rage from 86 kHz
  to 5 MHz.
Thus we concluded the photomultiplier, combined with
  the base used, met our requirements.

We note that the two of the photomultipliers,
 R3234 (the monitor photon detector) and R1398, 
 `see' quite different photons;
 the former sees mostly single photons with
 relatively low rates while the latter 
 sees much more intense photons with a rate up to 5 MHz.  
Thus an accidental cancellation of systematic errors is expected to be
 uncommon.
The result in turn gives good confidence to the monitoring method.
\section{Summary and Discussion}
In order to study photomultiplier's gain stability 
 at high counting rate, we constructed an LED pulsed light source and
 its output monitor system. 
For the monitor system, we employed 
 a photon counting method. 
It samples a small portion of light output and
 measures single photon rates with a monitor photon detector.
It thus monitors the relative light output from the source.
It is  virtually insensitive to the gain change 
 of a monitor photon detector because, 
 as long as the discrimination between the signal from
 background noise is clear, the rate of the single photon 
 count remains constant.

In our actual setup, we used  a photomultiplier
 as a monitor photon detector.
Thermal (random) noises and after-pulses were found to be 
 the two main backgrounds. 
We could make the errors due to these backgrounds 
 sufficiently small ($< 1 \% $)
 by selecting a suitable photomultiplier
 and a operating condition.
We tentatively assign $\sim 7\%$ as an absolute accuracy
 in this method.
This accuracy was estimated by the method described in \S 4.1.

Our direct application of this system was to investigate
 the gain stability of the photomultiplier (R1398)
 used in our trigger counter.
As shown in \S 4.2, it was proved that the
 photomultiplier and base system 
 could satisfy our requirements.
At the same time, it is found that the photon 
 counting method offers a simple way
 to monitor outputs from a pulsed light source.
Together with an LED light source, it provides a handy way to
 investigate photomultiplier's gain stability at high counting rates.
\begin{flushleft}
{\large {\bf Acknowledgments}}  \\
\end{flushleft}
It is our pleasure to thank Professors H. Sugawara, S. Yamada, 
S. Iwata, K. Nakai and K. Nakamura 
 for their support and encouragement.
We are grateful to Y. Higashi, S. Koike and T. Takatomi,
who are the member of Mechanical Engineering Center at KEK,
for their valuable help in making our trigger counters.
Y.T, Y.M and M.S acknowledge receipt of Research Fellowships of
 the Japan Society for the Promotion of Science
 for Young Scientists.


\begin{thebibliography}{99}
\bibitem{ref-photoelectron}
  In some photon detectors like a photomultiplier, we actually count
  photoelectrons instead of photons. We use these two words
  interchangeably when no confusion occurs.
\bibitem{note-LED-property}
 Since the number of tested LED types was limited, these observations
 may not hold for other blue and/or green LEDs.
\bibitem{ref-cutposition}
In order to illustrate the effect of the change in the cut position,
 we take a measurement described in \S 4.1 as an example.
The maximum change in the normalized photomultiplier gain
 (i.e. the gain rescaled to 1 at 0.7 MHz)
  was found to be less than 0.7\% 
  when the cut position was varied between 0.3 and 0.5.
\bibitem{note-Thermal-exception}
The exception was the cross-check measurement described in \S 4.1.
See also ref.~\cite{note-Thermal-systematic}.
\bibitem{note-ap-pmt}
  We expect that the after-pulse rate depends on, among other things,
   residual gas pressure of photomultiplier tubes.
  This means that the quantity $\alpha_{a.p}$ varies from one tube to another
   within one type of photomultipliers.
  (In general, old tubes have bigger $\alpha_{a.p}$ due to
    the increase of residual gas in tubes.)
  Since the history of the tested tubes was not quite known and 
   the number was limited (1 $\sim$ 3 tubes), 
   the selected photomultiplier type may not mean the best type in general.
  In principle, $\alpha_{a.p}$ may depend on 
    photomultiplier's gain (HV), too.
  However, very little dependence on the HV was found.
\bibitem{note-LED-drop}
As can be seen from the results, the light output from the LED 
  decreased with the repetition rate.
A part of the reason could be the long charge-up time constant
  for the LED capacitor $C_{LED}$ ($1/\tau \sim$ 9 MHz).
We left it as it was,
  because we thought it desirable to keep the average cathode
  current of R239 roughly constant independent of the pulse rate.
\bibitem{note-Thermal-systematic}
For this measurement, the smallest $\average{n}$ was 0.0001 (at 5 MHz).
Thus the thermal noise contribution could be as large as 2.4\%.
If this was the only source of discrepancy between the two sets
 of the measurement, the discrepancy should become
  larger as $\average{n}$ decreased.
We did not see such behavior; thus we concluded that the thermal
  noise was not, at least, the main source of the discrepancy.
\bibitem{note-7LEDs}
As can be seen from fig.\ref{fig-cathode} the LED light output was 
   roughly constant below 2 MHz.
\end{thebibliography}
\end{document}